\documentclass[twocolumn,english,aps,prl,superscriptaddress,amsmath,amssymb,showpacs]{revtex4-1}
\usepackage[T1]{fontenc}
\usepackage[latin9]{inputenc}
\usepackage{array}
\usepackage{multirow}
\usepackage{amsmath}
\usepackage{amssymb}
\usepackage{graphicx}

\makeatletter

\providecommand{\tabularnewline}{\\}

\usepackage{babel}

\makeatother

\usepackage{babel}
\begin{document}

\title{ESR study of atomic hydrogen and tritium in solid T$_{2}$ and T$_{2}$:H$_{2}$
matrices below 1K}

\author{S.Sheludiakov}

\affiliation{Department of Physics and Astronomy, University of Turku, 20014 Turku,
Finland }

\author{J.Ahokas}

\affiliation{Department of Physics and Astronomy, University of Turku, 20014 Turku,
Finland }

\author{J.Järvinen}

\affiliation{Department of Physics and Astronomy, University of Turku, 20014 Turku,
Finland }

\author{O.Vainio}

\affiliation{Department of Physics and Astronomy, University of Turku, 20014 Turku,
Finland }

\author{L.Lehtonen}

\affiliation{Department of Physics and Astronomy, University of Turku, 20014 Turku,
Finland }

\author{D.Zvezdov}

\affiliation{Department of Physics and Astronomy, University of Turku, 20014 Turku,
Finland }

\affiliation{Kazan Federal University, 18 Kremlyovskaya St., Kazan 42008, Republic
of Tatarstan, Russian Federation}

\author{S.Vasiliev}

\email{servas@utu.fi}

\affiliation{Department of Physics and Astronomy, University of Turku, 20014 Turku,
Finland }

\author{D.M. Lee}

\affiliation{Institute for Quantum Science and Engineering, Department of Physics
and Astronomy, Texas A\&M University, College Station, TX, 77843,
USA}

\author{V.V. Khmelenko}

\affiliation{Institute for Quantum Science and Engineering, Department of Physics
and Astronomy, Texas A\&M University, College Station, TX, 77843,
USA}

\date{\today}

\keywords{ESR}

\pacs{choose}
\begin{abstract}
We report on the first ESR study of atomic hydrogen and tritium stabilized
in a solid T$_{2}$ and T$_{2}$:H$_{2}$ matrices down to 70$\,$mK.
The concentrations of T atoms in pure T$_{2}$ approached $2\times10^{20}$cm$^{-3}$
and record-high concentrations of H atoms $\sim1\times10^{20}$cm$^{-3}$
were reached in T$_{2}$: H$_{2}$ solid mixtures where a fraction
of T atoms became converted into H due to the isotopic exchange reaction
$\mbox{T+H\ensuremath{_{2}\rightarrow}TH+H}$. The maximum concentrations
of unpaired T and H atoms was limited by their recombination which
becomes enforced by efficient atomic diffusion due to a presence of
a large number of vacancies and phonons generated in the matrices
by $\beta$-particles. Recombination also appeared in an explosive
manner both being stimulated and spontaneously in thick films where
sample cooling was insufficient. We suggest that the main mechanism
for H and T migration is physical diffusion related to tunneling or
hopping to vacant sites in contrast to isotopic chemical reactions
which govern diffusion of H and D atoms created in H$_{2}$ and D$_{2}$
matrices by other methods. 
\end{abstract}
\maketitle

\section{Introduction}

The solid hydrogens are among the simplest quantum crystals. Small
masses and weak interactions result in a dramatic influence of quantum
effects on their properties. Light impurities of the atomic hydrogens
introduced into such host matrices turned out to be mobile and able
to diffuse by the repetition of quantum exchange reactions \cite{Kumada2002}.
It might be expected that reducing the distance between unpaired hydrogen
atoms in the matrix by accumulating them to high concentrations may
lead to a number of fascinating phenomena related to quantum degeneracy
or to emergence of a strong exchange interaction between electron
clouds and possible conductivity of the matrix.

A number of methods are available for producing unpaired atoms inside
the molecular hydrogen solids. Among them are condensing of rf-discharge
products onto liquid-helium cooled surfaces\cite{Ivliev83} or directly
into superfluid helium\cite{Gordon83,Kiselev02}, $\gamma$-irradiation\cite{Miyazaki84}
or adding small amounts of $\beta-$radioactive tritium into an initial
gas mixture.  The highest densities of H in H$_{2}$ above 1~K have
been obtained with the latter method, $7\times10^{18}\,\textrm{cm}^{-3}$
with a 2\% T$_{2}$ admixture \cite{Collins93}, while densities approaching
$10^{20}\,\textrm{cm}^{-3}$ were reached in pure tritium \cite{Collins92}.
The method where $\beta$-decay of tritium produces free radicals
was pioneered by Lambe \cite{Lambe60} who studied radiation induced
defects in solid T$_{2}$ and by Sharnoff and Pound who studied accumulation
and dynamic nuclear polarization of D in  solid D$_{2}$ \cite{Sharnoff_Pound}.
 First experiments below 1~K before our work were carried out by
Webeler \cite{Webeler76}, who conducted a calorimetric study of H$_{2}$
with a 0.02\% tritium admixture  in a range of 0.2-0.8$\,$K. Both
spontaneous and stimulated heat spikes of the sample cell temperature
 were detected and attributed to collective recombination of H atoms
in the H$_{2}$ matrix. Collins et al. \cite{Collins90} repeated
Webeler's experiment using ESR as a detecting tool at a temperature
of 1.2$\,$K and confirmed that the heat spikes are also accompanied
by abrupt \emph{en masse} atomic recombination.  Mapoles et al. \cite{Mapoles90}
detected spectacular light emission in their D-T (50\% DT, 25\% D$_{2}$
and 25\% T$_{2}$) samples which was also assigned to explosive recombination
of atoms. The atomic concentrations achieved in the experiments of
Webeler were estimated indirectly as $5\times10^{17}$cm$^{-3}$,
while in following theoretical works, Zeleznik\cite{Zeleznik76} and
Rosen\cite{Rosen76} hypothesized that they could be significantly
increased if the storage temperature would be lowered. These theoretical
conclusions were supported by Collins et al. who used X-band ESR for
studying deuterium and tritium atoms in solid D-T mixtures at temperatures
2.1-10$\,$K and found a strong dependence of the steady state concentration
of atoms on storage temperature \cite{Collins92}.

Another method for producing high concentrations of H atoms was recently
employed by Ahokas et al.\cite{HinH2Rapid09}, \cite{Ahokas10} who
used a cryogenic rf discharge to dissociate H$_{2}$ molecules in
solid films $in$ $situ$ and reached H concentrations $\simeq2\times10^{19}$cm$^{-3}$
at 0.5~K. Later it was demonstrated that even higher concentrations
of atomic hydrogen can be produced in H$_{2}$:D$_{2}$ mixtures where
deuterium atoms become converted into H in the course of the isotopic
exchange reactions $\mbox{D+H\ensuremath{_{2}\rightarrow\mbox{HD+H}}}$
and $\mbox{D+HD\ensuremath{\rightarrow}\ensuremath{\mbox{D\ensuremath{_{2}+\mbox{H}}}}}$\cite{DNPPRL14}.

Similar to the lighter counterparts, the isotopic exchange reactions
of hydrogen and deuterium with tritium (\ref{eq:faster}) and (\ref{eq:slower})
should proceed in T$_{2}$:H$_{2}$(D$_{2}$) matrices.

\begin{equation}
\mbox{T\ensuremath{+\mbox{H\ensuremath{_{2}\rightarrow\mbox{TH\ensuremath{+\mbox{H}}}}}}}\label{eq:faster}
\end{equation}
\begin{equation}
\mbox{T+TH\ensuremath{\rightarrow\mbox{T\ensuremath{_{2}+\mbox{H}}}}}\label{eq:slower}
\end{equation}

The rates of these reactions in the gas phase were calculated by Truhlar
et al. \cite{Truhlar83} and later by Aratono et al. \cite{Aratono00}
who predicted an extremely high rate of the reaction (\ref{eq:faster})
at low temperatures, $k^{ex}\sim2\times10^{-25}$cm$^{3}$s$^{-1}$,
and a two order of magnitude smaller rate for the reaction (\ref{eq:slower}).
The only experimental observation of the isotopic exchange reactions
(\ref{eq:faster}) and (\ref{eq:slower}) at low temperatures so far
was done by Aratono et al. \cite{Aratono98} who studied them in superfluid
helium where they produced T atoms from $^{3}$He $in$ $situ$ by
neutron bombardment.  The authors were unable to deduce  the absolute
reaction rates, but reported on the isotope effect when H is replaced
with D in the reactions \eqref{eq:faster} and \eqref{eq:slower}.
For the reaction \eqref{eq:faster} it was measured to be $k_{\textrm{H}_{2}}^{ex}/k_{\textrm{D}_{2}}^{ex}\approx150$,
while for reaction \eqref{eq:slower} $k_{\textrm{TH}}^{ex}/k_{\textrm{TD}}^{ex}<19.6.$
Reactions similar to (\ref{eq:faster}) and (\ref{eq:slower}) involving
deuterium and tritium atoms should proceed much slower than those
for hydrogen because of smaller zero-point energies of the reactants.
This was also supported by the only experimental study of D and T
atoms in a D-T matrix carried out by Collins et al. \cite{Collins92}
at T=2.1$\,$K who reported on the abscence of T-to-D conversion in
their samples. The authors estimated the total atomic concentrations
to be $\sim$10$^{20}$cm$^{-3}$ with a possible uncertainty $\sim$50\%.
Studying the mixtures of T and H in this work we found that the isotopic
exchange reaction (\ref{eq:faster}) proceeds with a much higher rate
than reaction (\ref{eq:slower}). We estimate the rate of the reaction
(\ref{eq:faster}) to be $k=3(2)\times10^{-26}$ cm$^{3}$s$^{-1}$.

The main purpose of these experiments was to examine opportunities
for reaching the highest possible concentrations of atomic species
using the $\beta$-decay of tritium. We performed the first quantitative
ESR study of T and H atoms stabilized in thin tritium films at temperatures
below 1$\,$K down to 70$\,$mK. The record-high concentrations of
T atoms in pure T$_{2}$ approaching $2\times10^{20}$cm$^{-3}$ and
concentrations of H atoms of about $1\times10^{20}$cm$^{-3}$ in
solid T$_{2}:$H$_{2}$ were reached. It turned out that in the films
thicker than 100$\,$nm, maximum achievable density of H and T was
limited by a spontaneous and stimulated explosive recombination of
atomic species. The explosion threshold density and periodicity turned
out to be strongly dependent on the storage temperature. Decreasing
the film thickness below 100$\,$nm allowed us to avoid explosions,
but the maximum density was also somewhat reduced due to less effective
use of the electrons for dissociation of molecules.

\section{Experimental}

\subsection{Setup}

\begin{figure}
\includegraphics[width=1\columnwidth]{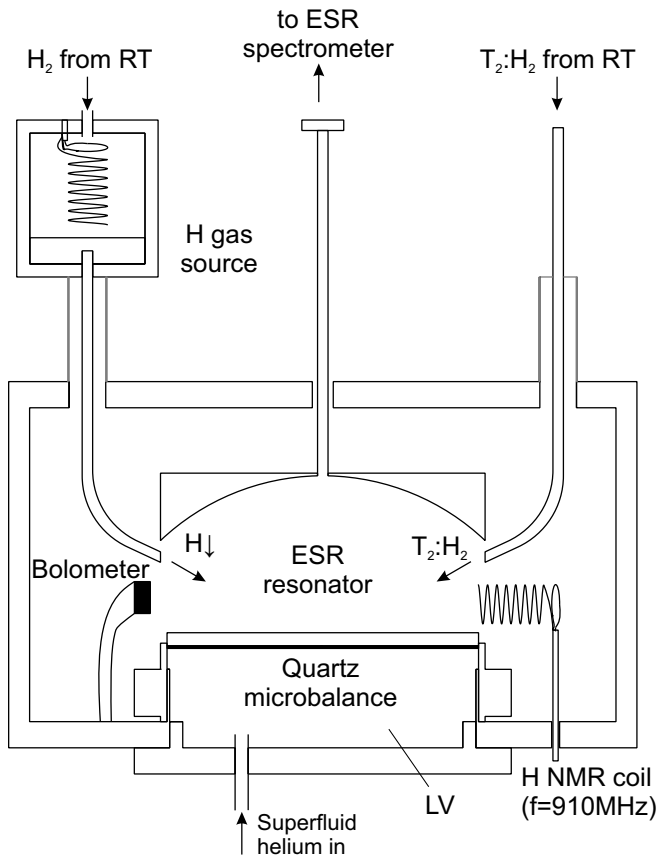}

\protect\protect\caption{Sample cell schematic.The T and D NMR coils are not shown in the figure\label{fig:Sample-cell}}
\end{figure}

The experiments were performed in the sample cell (SC) shown in Fig.\ref{fig:Sample-cell}
with further details described in \cite{Cellpaper}. The SC is located
at the center of a 4.6$\,$T superconducting magnet and is attached
to the mixing chamber of an Oxford 2000 dilution refrigerator. The
main investigation tools in our experiments are a 128$\,$GHz ESR
spectrometer and a quartz-crystal microbalance (QM) able to measure
the film thickness with a 0.2 monolayer accuracy. The ESR resonator
(Q=5700 at 300$\,$mK) has an open Fabry-Perot geomentry which made
it possible to install beam lines for condensing films of hydrogen
isotopes onto the QM and provide a flux of atomic hydrogen created
in a specially constructed H-gas source. Three auxillary rf resonators
for performing electron-nuclear double resonance (ENDOR) of H, D and
T atoms are arranged near the QM. Only one of them, the H NMR coil,
is shown in Fig.\ref{fig:Sample-cell}. A capillary for condensing
the molecular hydrogen comes directly from the room temperature gas
handling system and it is kept above the tritium boiling temperature
during condensing of the film by driving current through electrical
heaters. A special source of cold hydrogen gas is arranged at the
top of the sample cell body and connected to it via a stainless steel
tubing system. The gas of H atoms is very useful for calibration of
the absolute number of spins detected by our ESR spectrometer as well
as for the accurate measurement of the magnetic field and the ESR
line shifts \cite{Cellpaper}. The source can be also used for obtaining
a flux of molecular hydrogen, which was not used in the present work.

A Ru-oxide bolometer is arranged in close proximity to the sample
to measure the heat released during explosive recombination. The bolometer
was suspended on fine superconducting wires which assure only a weak
thermal coupling to the sample cell body. The bolometer has a negligibly
small heat capacity and even tiny amounts of heat can quickly raise
its temperature above the temperature of the sample cell.

\subsection{Procedure}

The technique of condensing hydrogen onto a cold surface via the long
and sufficiently cold capillary provides a rather efficient way of
cleaning the condensed gas from any other contaminants except hydrogen
and its isotopes. Therefore, for the experiments described in this
work, we have not paid much attention to the chemical purity of T$_{2}$
gas, and utilized the most simple and cheap source available. The
T$_{2}$ gas we used was extracted from commercially available tritium
vials produced for use as luminescent fishing floats. Each such 5
mm diameter and 5 cm long vial contained about 5-10 $\mu$moles of
T$_{2}$, which was sufficient for several experiments described in
this work. The purity of the gas extracted from the vials was verified
using a MKS-Granville-Phillips VQM 835 mass spectrometer with the
main concern being the amount of other hydrogen isotopes in it. It
turned out that as the main impurity, the gas in the vials typically
contained about 15\% of HT. It is known that the main impurity in
the T$_{2}$ gas right after production should be DT($\sim1\%$),
while significant HT contamination appears during storage\cite{souers1986hydrogen}.
We found that T$_{2}$ gas quickly degraded after extracting it from
the glass vials. About 20\% of it became converted to HT and H$_{2}$
after 3 months of storage. In order to minimize the HT content we
used only fresh T$_{2}$ gas for preparing our samples.

The films of solid molecular tritium and hydrogen were deposited by
condensing a few $\mu$mol of normal T$_{2}$ or T$_{2}$:H$_{2}$
mixtures onto the QM directly from a room temperature reservoir. A
small amount of helium ($\sim1\,$mmol) was condensed into the
lower volume (LV in Fig.\ref{fig:Sample-cell}) under the quartz microbalance
disk in order to have a saturated helium film there. The film flushes
the lower surface of the QM and provides an efficient way to remove
an excess of heat released during film deposition, recombination of
atoms and tritium $\beta$-decay. The SC temperature during sample
deposition stayed between 1 and 1.5$\,$K.

Unpaired H and T atoms appeared in the solid films, and were detected
by the ESR spectrometer immediately after completing the deposition
process. The atoms result from the dissociation of molecules by the
5.7$\,$keV electrons of the T$_{2}$ $\beta$-decay. We followed
the kinetics of the growth of the atomic concentrations during the
time interval from several days up to two weeks, periodically recording
the ESR spectra and following the QM frequency changes. The SC temperature
was stabilized to several different values below 1$\,$K in order
to study the sample properties as a function of temperature. It turned
out that the presence of tritium introduces substantial heating in
the SC and limits minimum attainable temperature to about 150$\,$mK
for the films with thickness about 300$\,$nm. The sample cell temperature
was also shortly raised by several tens of mK during the magnetic
field sweeps due to eddy current heating. This was essential for triggering
the explosive recombination in some of the samples. After finishing
the measurements for each film, the sample cell and filling capillary
were heated to a temperature above 30$\,$K and pumped for several
hours in order to evacuate all hydrogen and clean all surfaces before
creating a new sample. This cycle was repeated several times for samples
of different H$_{2}$:T$_{2}$ composition and thickness.

For calibration purposes we utilized the ability to accumulate the
H gas in the main volume of the SC. This is normally done by: 1) condensing
a certain amount of molecular hydrogen into the H gas source; 2) condensing
small amounts of helium inside the upper volume of the SC and the
source, sufficient to form a several nm thick superfluid $^{4}$He
film ; 3) running the rf discharge in the miniature rf coil inside
the H gas source. The gas of atomic hydrogen is produced by dissociation
by the electrons of the discharge. The helium film covering the surfaces
prevents recombination of the atoms on the SC walls. In the course
of the experiments we found that atomic hydrogen gas can be accumulated
(with a substantially smaller rate) also without running discharge,
once a helium film is present in the SC. This effect was only observed
in the experiments with tritium-hydrogen mixtures. It indicates that
a fraction of the atoms resulting from H$_{2}$ dissociation in solid
films by the electrons of the $\beta$-decay may be kicked out from
the films into the bulk of the SC and get accumulated there. Unfortunately
we were unable to observe gas lines of spin-polarized tritium. This
is caused by substantially larger adsorption energy for T on helium
surfaces and possibility that atomic T may penetrate the helium film.
Presence of a helium film in the SC leads to an extra heat load caused
by the  film re-condensing from the upper parts of capillaries. In
order to avoid this disturbance and reach the lowest temperatures
most of the experiments described in this work were performed without
helium film in the SC.

\subsection{Samples}

The main emphasis of this work was placed on reaching the highest-possible
concentration of unpaired atoms in hydrogen solids using $\beta$-decay
of tritium. Aiming on that we tried to find optimal values of two
basic parameters of the films: the film thickness and T$_{2}$:H$_{2}$
ratio. We studied both pure tritium films and different mixtures of
T$_{2}$ and H$_{2}$. We expected that in the latter a significant
fraction of T atoms will be converted to H by the chemical exchange
reactions (\ref{eq:faster}) and (\ref{eq:slower}). Varying the film
thickness we tried to reach a trade-off between more efficient generation
of unpaired atoms in thicker films and better cooling expected for
thinner samples. A summary of the properties of 6 different samples
studied in this work is given in Table \ref{tab:A-summary-table}.

A second goal of this work was to study in detail explosive recombination
of H and T atoms previously observed in works of Webeler \cite{Webeler76}
and Collins et al. \cite{Collins90}. Varying the film parameters:
thickness and isotope composition as well as sample temperature, we
tried to find the conditions for which this process can be suppressed.

First a 1$\,\mu$m thick sample of para-H$_{2}$ with a 1\% tritium
admixture was studied at T=150$\,$mK. No signals of atomic tritium
were observed while the H concentrations obtained in this sample levelled
off at about 1$\times10^{19}$cm$^{-3}$. We did not detect any signatures
of explosive recombination of atoms similar to what were reported
in \cite{Webeler76} and \cite{Collins90} for the bulk samples. This
can be explained by a larger surface to volume ratio and much better
cooling of our hydrogen films compared to previous experiments. In
the work \cite{Collins90} authors reported on suppression of events
of spontaneous explosive recombination by providing better cooling
to their samples after collecting bulk amounts of liquid helium in
the sample cell. The films we studied are rather thin compared to
the penetration depth of electrons ($\sim3.5\mbox{\,}\mu$m) released
in $\beta$-decay of tritium \cite{Schou78} and only a fraction of
their kinetic energy, $\langle E_{k}\rangle=$5.7$\,$keV, is dissipated
in the film. This reduces the heat released in the samples, but also
reduces the rate of dissociation in thin films leading to smaller
atomic densities.

Next, we studied films of \textquotedbl{}pure\textquotedbl{} T$_{2}$,
i.e. without any H$_{2}$ added to it prior to condensing. A second
sample was a 250$\,$nm pure tritium film, which we studied at the
lowest temperature of 160$\,$mK, limited by the heat from tritium
decay. In this sample we reached maximum atomic concentrations approaching
$2\times10^{20}$cm$^{-3}$ where most of the atoms were T, with the
T:H ratio being about 6:1.

The atomic concentrations in this sample were limited by periodic
collective recombination of atoms which raised the SC temperature
from 0.16K to about 0.25K. Similar heat spikes were also registered
by the bolometer, which has a much faster response time. The spikes
were found to appear both while sweeping magnetic field and between
the sweeps when the cell was gradually cooling down. Collective recombination
also resulted in a partial sublimation of tritium films detected by
the QM. A fraction of evaporated tritium molecules re-condensed onto
the other surfaces of the SC including the spherical mirror.

A different behaviour was observed in a substantially thinner, 35$\,$nm
pure T$_{2}$ film (sample 3). The accumulation of atoms in the film
was not interrupted by their explosive recombination. Condensing a
smaller amount of tritium also allowed us to store the sample at temperature
of about 80$\,$mK. However, the maximum densities of atoms in the
thinner film were a factor of 2 lower than that in Sample 2.

Three samples with different T$_{2}$:H$_{2}$ content were studied
to examine the possibility of reaching the highest concentrations
of H atoms and possible observation of the exchange reactions between
two isotopes. Sample 4: 80$\,$nm T$_{2}$:4\%H$_{2}$, sample 5:
T$_{2}$:30\%H$_{2}$ (300$\,$nm) and sample 6: H$_{2}$:5\%T$_{2}$
(360$\,$nm). Total density of atoms was a factor of 2 lower than
in thick \textquotedbl{}pure\textquotedbl{} tritium sample, but in
Sample 5 we succeeded in reaching a record high density of H atoms
exceeding $10^{20}$cm$^{-3}$. We observed no explosive atomic recombination
in the samples of the T$_{2}$:H$_{2}$ mixture films. All these samples
featured a much smaller T:H ratio as compared to the initial content
of T$_{2}$:H$_{2}$ which we interpret in terms of the isotope exchange
reaction.

\begin{table}
\begin{tabular*}{1\columnwidth}{@{\extracolsep{\fill}}@{\extracolsep{\fill}}|c|c|c|c|c|c|c|c|c|c|}
\hline 
\multirow{2}{*}{ } & \multirow{2}{*}{$s$, nm} & \multicolumn{3}{c|}{Composition,\% } & \multicolumn{3}{c|}{$n\times$10$^{19}$ cm$^{-3}$} & \multicolumn{2}{c|}{H/(T+H)}\tabularnewline
\cline{3-10} 
 &  & T$_{2}$  & H$_{2}$  & HT  & T  & H  & total  & observed  & expected\tabularnewline
\hline 
\hline 
1.  & 1000  & 1  & 99  & -  & -  & 1.0  & 1.0  & 1.00  & 0.99\tabularnewline
\hline 
2.  & 250  & 85  & -  & 15  & 15  & 2.5  & 18  & 0.14  & 0.08\tabularnewline
\hline 
3.  & 35  & 85  & -  & 15  & 10  & 1.0  & 11  & 0.09  & 0.08\tabularnewline
\hline 
4.  & 80  & 81  & 4  & 14  & 4.2  & 4.5  & 8.7  & 0.53  & 0.11\tabularnewline
\hline 
5.  & 300  & 60  & 29  & 11  & 0.5  & 10.5  & 11  & 0.95  & 0.35\tabularnewline
\hline 
6.  & 250  & 4  & 95  & 1  & -  & 2.5  & 2.5  & 1.00  & 0.96\tabularnewline
\hline 
\end{tabular*}

\protect\protect\caption{A summary table of the samples studied in this work. Film thickness
is denoted as $s$. Expected values of H/(T+H) ratios were estimated
from H$_{2}$ and HT content.\label{tab:A-summary-table} }
\end{table}

\subsection{ESR and ENDOR spectra}

ESR spectra were recorded by a cryogenic heterodyne spectrometer which
does not utilize field or frequency modulation \cite{Vasilyev04}.
In this work we used the CW method of operation, where the frequency
of the excitation source is kept constant, while the magnetic field
is swept across the resonance. The ESR signal at the output of the
detection system contains both components of the complex magnetic
susceptibility: absorption and dispersion as a function of the magnetic
field sweep. In all our spectra presented below we will show absorption
signals only.

\begin{figure}
\includegraphics{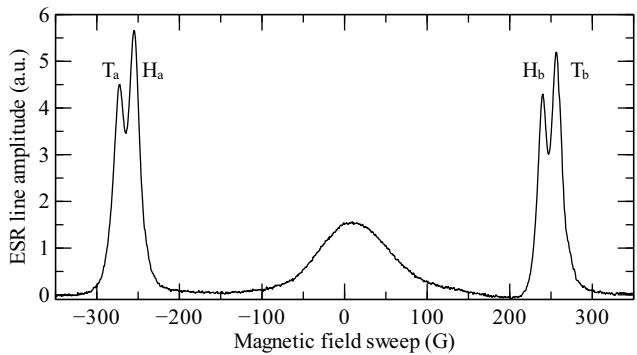}

\protect\protect\caption{A panorama spectrum of T and H ESR lines of sample 4 stored at 70$\,$mK.
Note the difference in T and H line polarization. A broad line of
unknown origin appears at the ESR spectrum center\label{fig:A-panorama-spectrum}}
\end{figure}

The ESR spectrum of atomic hydrogen and tritium contains a doublet
of lines separated by a distance equal to the hyperfine interaction.
The hyperfine splitting between the two lines is $\approx507\,$G
for H and slightly larger $\approx541\,$G for T. A typical ESR  spectrum
which includes all lines of atomic hydrogen and tritium in T$_{2}$:H$_{2}$
matrices is shown in Fig.\ref{fig:A-panorama-spectrum}. The spectrum
shown corresponds to Sample 4, and is taken after the accumulation
of atomic species has been saturated and a maximum density of atomic
species is reached. The T:H line ratio was different for various samples.
T lines in the samples 1 and 6 were not detected at all, while H lines
in the samples 2 and 3 were substantially weaker than T lines. Even
though the difference in the hyperfine constants looks relatively
large, for the very high densities studied in this work the density
dependent broadening was so large that it was not always easy to resolve
ESR line of H and T from each other.

At high densities of atoms studied in this work the effects of dipole-dipole
interaction between atoms start to play important role and influence
the shape and position of ESR lines. It has been shown in our previous
work \cite{Ahokas10} that the dipole-dipole interaction leads to
a Lorentzian lineshape with the width linearly increasing as a function
of density. This dependence can be used for the absolute determination
of atomic density. A second important consequence of high density
fully polarized electron spins is a macroscopic magnetization of the
sample, which leads to shifts of the ESR lines. This effect depends
on the geometry of the sample, and for thin films perpendicular to
the main polarizing field the net dipolar field is opposite to the
main polarizing field. This leads to the linear density-dependent
shift of the ESR lines towards the larger sweep fields, i.e. to the
right in the spectra as they are recorded by our technique. Having
the possibility of using the H gas lines as absolute field markers,
these shifts can be accurately measured and also provide a measure
of the absolute density of atoms in the films.

For the high density samples and thick (>350$\,$nm) films another
ESR line broadening effect caused by the radiation damping brings
extra complications in the analysis of the ESR spectra. The effect
of radiation damping is related to spontaneous and coherent emissions
of energy stored in the spin system into the resonant cavity at the
electron Larmor frequency \cite{Bloembergen54}. It typically occurs
in spin systems strongly coupled to the microwave field of the cavity,
when the relaxation time due to interaction of spin system with cavity
field becomes comparable to the spin-spin relaxation time $T_{2}$,
i.e. $T_{2}\sim T_{rad}=(2\pi M_{0}Q\eta)^{-1}$. Here $M_{0}$
is the sample magnetization, $Q$ is the resonator quality factor
and $\eta$ is the resonator filling factor. In this case the effective
relaxation time of the system $1/T_{2}^{'}$=$1/T_{2}$+$1/T_{rad}$.
This also results in an additional line broadening proportional to
the number of spins in the sample. The radiation damping effect depends
on the detuning of the spectrometer frequency from the center of the
cavity resonance, and it is possible to reduce it to a negligible
level by increasing the detuning to several cavity resonance widths.
This was verified for the thickest films of the sample 1, where the
effect was strongest. We observed a factor of 2 decrease in the ESR
linewidth after such detuning. For all other samples the radiation
damping effects were substantially weaker and added not more than
20\% to the actual linewidth, which was however taken into account
in the analysis.

\begin{figure}
\includegraphics[width=1\columnwidth]{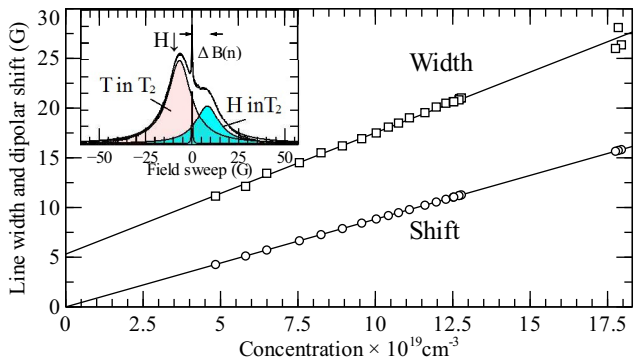}

\protect\protect\caption{The ESR line width and shift from the H-gas phase line as a function
of the atomic concentration. Inset: a typical spectrum of the H and
T low field lines fitted by 3 lorentzian curves. The concentration-dependent
(dipolar) shift from the H-gas line (H$\downarrow$) is labeled as
$\Delta$B(n) in the inset\label{fig:Shifts}. }
\end{figure}

A typical dependence of the tritium ESR line shift and width on the
total atomic density measured for the sample 2 is presented in Fig.\ref{fig:Shifts}.
In the inset we show a real spectrum of the high field line with all
three components: T and H lines from the atoms in the film, and the
H gas line from the atoms in the bulk of the SC. In this sample the
lines from the atoms in the solid are of comparable amplitude and
poorly resolved from each other because of their rather large width.
The H gas line is very narrow and easily distinguishable in the spectrum.
We fitted such lineshapes with three Lorentzian functions, which are
also presented in the inset of the Fig.\ref{fig:Shifts}.

For the measurement of the absolute density of atoms in thin solid
films we used known dependence of the ESR line width and shift on
density \cite{Ahokas10}. This dependence was verified during several
experimental runs with respect to the calorimetric method of measurement
of the absolute number of hydrogen atoms. The calorimetric method
is based on the measurement of the energy released in stimulated by
ESR recombination of the atoms of the H gas versus the reduction of
their ESR signal. This method agreed to within 20\% accuracy with
calibrations based on the linewidth and shift measurements, and we
consider 20\% as an upper limit estimate of the error in density determination
in this work.

A broad, $\sim100$ $\,$G wide, line of an unknown nature was observed
at the center of the ESR spectrum for all samples studied in this
work (Fig.\ref{fig:A-panorama-spectrum}). The line width and area
increased during storage and saturated after one week of measurements.
The line remained in the spectrum after the SC cleaning procedure
when we raised the SC temperatures above $\simeq20$ $\,$K and cooled
back to 1 K. Since the solid molecular film has been completely removed
after such cleaning, it seems that the central line originates from
some atoms or free radicals in the metallic mirrors of the Fabry-Perot
resonator. We have not observed this line after warming up to room
temperature and starting new experimental run. A large area of the
central line also implies a large number of the free radicals which
produce it, about same as the number of unpaired atoms in our molecular
films. A strong broad signal seen at the center of the spectrum in
Fig.\ref{fig:A-panorama-spectrum} is reminiscent of the central peak
seen in the semiconductor Si:P at high phosphorus concentration and
was taken to be attributed to the formation of the donor pairs coupled
by the strong exchange interaction \cite{Slichter55}. The work is
continuing in an attempt to fully understand this important observation.

A weak line of trapped electrons similar to that found for other hydrogen
matrices \cite{Electrons_QFS} was observed in thick T$_{2}$ and
T$_{2}$:H$_{2}$ samples but it was absent in the films thinner than
100$\,$nm (samples 3 and 4). In our ESR spectra we have not observed
any signatures of ions or other species trapped in hydrogen films.
The most stable of ions H$^{+}$ and H$_{3}^{+}$(T$^{+}$ and T$_{3}^{+}$)
have a zero electron spin and cannot be detected by ESR. The yield
of other ions, such as H$_{2}^{+}$ (T$_{2}^{+}$) and H$_{2}^{-}$
(T$_{2}^{-}$) is expected to be 4 orders of magnitude smaller than
for unpaired atoms\cite{MiyazakiBook04} and lies below our sensitivity
threshold.

In addition to the conventional ESR diagnostics we implemented measurements
of the Electron-Nuclear Double Resonance (ENDOR) on our samples. The
method is based on detecting the frequency of the NMR transition by
its influence on the amplitude of the ESR lines. The ENDOR measurement
was typically done by sweeping the magnetic field to the center of
the ESR line, and then applying the rf excitation to one of the miniature
coils located near the sample on the QM electrode. Then the frequency
of the rf source was swept slowly near the expected NMR transition
of H or T atoms. The NMR frequencies can be determined as

\begin{equation}
\omega_{NMR}\simeq2\pi\frac{A}{2}+\gamma B\label{eq:ENDOR formula}
\end{equation}

where $A$ is the hyperfine constant of H (1417.3~MHz) or T (1512.6~MHz),
$\gamma$ is the proton or triton gyromagnetic ratio and $B$ is the
local magnetic field felt by atoms. A change of the ESR signal was
observed when the frequency of the rf source matched the NMR transition.
Typical ENDOR spectra recorded by this method are presented in Figs.\ref{fig:ENDOR_TinT2}
and \ref{fig:ENDORs-of-H}.

\section{Experimental results }

\subsection{Pure tritium samples}

In our \textquotedbl{}pure\textquotedbl{} tritium samples we have
not added any extra hydrogen to the tritium gas which we extracted
from the vials. Since the analysis showed a 15\% HT impurity, we assume
that all our ``pure'' tritium films contained about 15\% of HT.
This is the smallest hydrogen impurity which could be realized in
experiments with tritium, but it is still rather large to assure abscence
of H atoms in the samples. Our study includes two pure tritium samples,
2 and 3 (Tab.\ref{tab:A-summary-table}) with thickness of 250 and
35 nm respectively.

\begin{figure}
\includegraphics{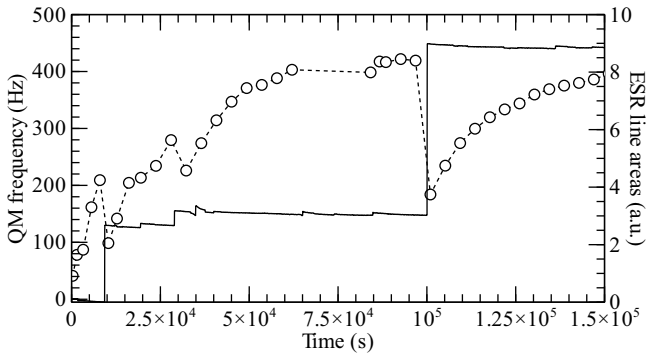}\protect\protect\caption{Accumulation of atomic concentration in the sample 2 (circles), and
QM response on explosive recombination of atoms in sample 2. \label{fig:QM-response-on}}
\end{figure}

\begin{figure}
\includegraphics[width=1\columnwidth]{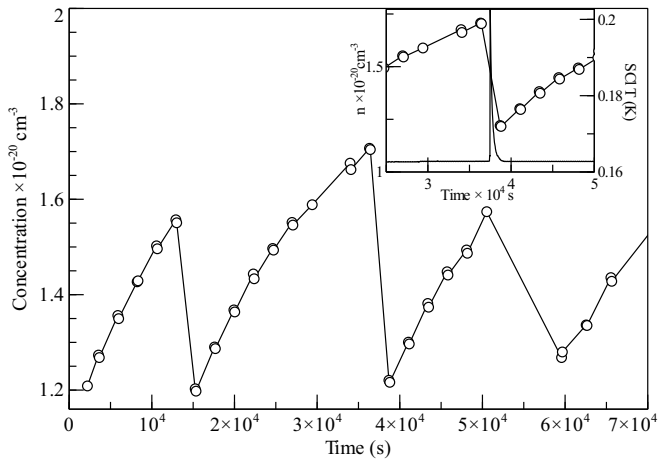}\protect\protect\caption{Time evolution of the total T and H atomic concentration in sample
2 stored at 160$\,$mK \label{fig:Explosions}. The explosive recombination
events presented here appear spontaneously. Inset: the sample cell
temperature spike during one of such events. }
\end{figure}

Accumulation of T atoms due to energetic electrons from T-decay in
sample 2 recorded at 160$\,$mK is shown in Figs. \ref{fig:QM-response-on}
and \ref{fig:Explosions}. Tritium ESR lines appeared almost immediately
after film deposition with a width of $\sim$5$\,$G. Then, the lines
grew rapidly, and already after about 3 hours we observed an abrupt
decrease in the accumulated density accompanied by the increase in
the QM frequency. Further accumulation was interrupted several times
with similar events at 3$\times10^{4}$ and a substantially larger
change at around $10^{5}$ sec on the time scale of Fig. \ref{fig:QM-response-on}.
These changes in the ESR and QM signals were accompanied by spikes
in the SC temperature up to $\sim$250 mK, and indicate explosive
recombination of part of the sample. Later on the explosive events
occurred for nearly equal time intervals, and with reproducible changes
of the atomic density from $\approx1.6\times10^{20}$ cm$^{-3}$ to
$\approx1.2\times10^{20}$ cm$^{-3}$. Occasionally the explosions
were triggered by the SC overheating during sweeps of magnetic field
between ESR lines. The T:H ratio did not change after the explosions.

Accumulation of the atomic densities in the much thinner (35 nm) Sample
3 followed monotonically increasing function which saturated at a
total density of $\approx1.2\times10^{20}$ cm$^{-3}$ and remained
stable at temperatures down to 70$\,$mK for the observation time
of several days. No explosive recombination events were seen for Sample
3. Smaller concentrations in the thinner film can be explained by
less efficient accumulation of atoms because the high-energy electrons
escape from the film carrying substantial part of their energy, which
is not used for dissociation.

Signals of atomic hydrogen behaved similarly to those of T in both
samples, but the total area of H ESR lines was 6 to 10 times smaller
than that for the T lines. This roughly corresponds to the T:H atomic
ratio in the gas mixture prior to condensing and did not change during
the course of measurement. This means that the majority of H atoms
in these samples were created by direct dissociation of HT molecules
while the isotopic exchange reaction $\mbox{T+HT\ensuremath{\rightarrow}T\ensuremath{_{2}}+H}$
is quite inefficient and does not lead to significant T-H conversion.

A zero-concentration or matrix width of $\approx5\,$G for H and T
lines in pure normal T$_{2}$ can be found as an offset on the vertical
axis of the width vs. density plot (see Fig.\ref{fig:Shifts}). This
value is substantially larger than that in n-H$_{2}$ (1.1$\,$G).
The line broadening due to ortho-T$_{2}$ molecules should rapidly
vanish due to the fast ortho-para conversion which is catalysed by
high concentration of atoms. The conversion should proceed for the
molecules in the closest neighbourhood of T and H atoms with a time
constant of about 10$\,$s and to be of order of $\simeq$10$^{3}$s
for the whole sample \cite{Gaines79}. We suggest that the main contribution
to the matrix width of H and T in our T$_{2}$ samples comes from
a large HT impurity. A HT molecule is composed of distinguishable
atoms and the magnetic moments of a proton and triton may contribute
to the line broadening. The matrix width for unpaired atoms in HD,
2.8$\,$G is known to be larger than that in n-H$_{2}$ (1.1$\,$G)
and o-D$_{2}$(1.3$\,$G). Taking into account that the deuteron magnetic
moment is about 3.5 times smaller than that of T, one may expect a
nearly 2 time larger matrix width in pure HT than in HD and some fraction
of that in our samples.

\begin{figure}
\includegraphics{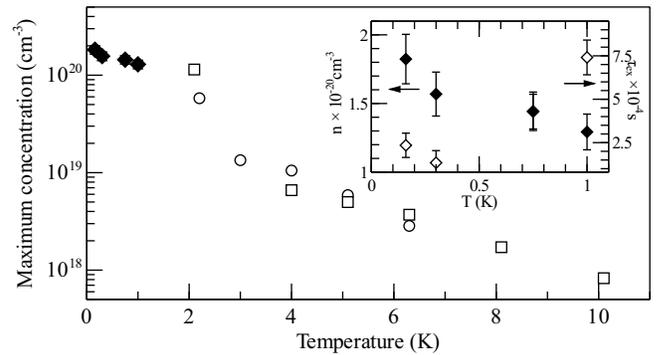} \protect\protect\caption{Maximum concentrations of unpaired atoms obtained in our work and
in previous works using tritium disintegration. Present work (sample
2) T in T$_{2}$ (black diamonds), T in T$_{2}$\cite{Collins92}
(open squares), D, T in D-T matrix (50\% DT, 25\% D$_{2}$ and 25\%
T$_{2}$) \cite{Collins92} (open circles) Inset: Dependence of atomic
concentration on storage temperature (sample 2) black diamonds and
dependence of average period of explosive recombination events on
storage temperature: open diamonds. The inset data was measured in
sample 2 \label{fig:Explosion_comp}}
\end{figure}

Recombination explosions of unpaired atoms were studied at different
temperatures: 300, 750$\,$mK and 1$\,$K. (see Fig.\ref{fig:Explosion_comp}).
Increasing the temperature slowed down growth of density as the recombination
rate increased. The heat spikes were observed at all three temperatures,
but the time intervals between the spikes greatly increased and changed
from about $2\times10^{4}$ $\,$s at 160 and 300$\,$mK to $8\times10^{4}$s
at 1K. This trend agrees with the results obtained previously by Collins
et al.\cite{Collins90,Collins92}who observed the heat spikes in their
bulk H$_{2}$+2\% T$_{2}$ samples at 1.2$\,$K and reported on their
absence at 2$\,$K. Storing the sample at higher temperatures also
resulted in smaller atomic concentrations reached. Similar results
were also found by Collins et al.\cite{Collins92} who reported on
gradual decrease of atomic concentrations while raising the storage
temperature from 2.1 to 10$\,$K.

\begin{figure}
\includegraphics{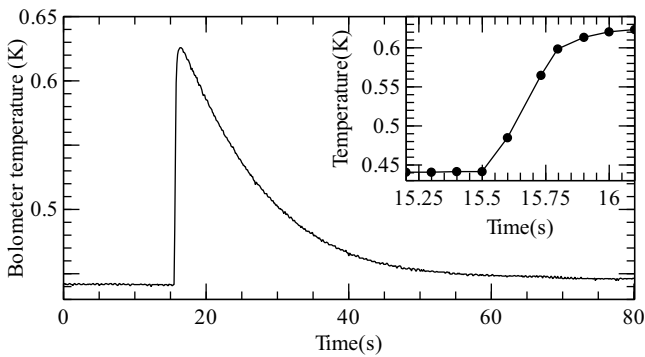}\protect\protect\caption{Bolometer response on the explosive recombination of atoms in sample
2. The bolometer temperature jump due to thermal explosion is zoomed
in the inset\label{fig:Bolometer-responce}}
\end{figure}

We studied the onset of thermal explosions using a Ru-oxide bolometer
which was arranged in the upper volume of the SC near the QM surface
with the samples. The bolometer temperature was somewhat higher than
the SC temperature due to heating by excitation during measurement
and noise pick-up (Fig.\ref{fig:Bolometer-responce}). The bolometer
temperature increased significantly during explosive recombination
from 0.4$\,$K to almost 0.65$\,$K which is much higher than the
cell temperature at the same time. There are two possible ways of
heating the bolometer: condensation of hot tritium and hydrogen molecules
onto it or by means of radiation similar to that observed by Mapoles
et al.\cite{Mapoles90}. The bolometer temperature after each event
of explosive recombination recovered to slightly higher reading which
brings evidence of re-condensing T$_{2}$ onto it.

Although the thermal response of the bolometer to the heat spikes
is substantially faster than that of the sample cell, neither of them
is fast enough to follow the actual time evolution of the recombination
explosions. From the bolometer response we can conclude that the explosion
event develops in the time scale faster than $\sim$0.1 s. Mapoles
et al. detected flashes in bulk D-T samples which were attributed
to explosive recombination of atoms\cite{Mapoles90}. The flashes
they observed developed within $\leq$1$\,$ms which can be assumed
as an upper limit for formation of a heat spike. We were not able
to resolve such fast events with the technique available in these
experiments.

We tried to measure the energy released in the explosions by simulating
a heat spike with heat pulses applied to the resistive heater attached
to the sample cell body. This was done by adjusting the strength and
duration of the heat pulses to obtain the same SC thermometer response
as was detected during explosive recombination. We observed the same
response using a 17$\,$ms and $P\simeq22\,$mW pulse. Then, calculating
the total energy released by such pulse we evaluated the number of
recombined atoms which would produce such energy bursts as 1.0$\times10^{15}$.
This number matches well the decrease in atom number 9.0$\times10^{14}$
measured by our ESR technique. A good agreement between these two
numbers also proves good accuracy of our calibration of the absolute
density as a function the ESR line width and area.

Accumulation of unpaired atoms in solid T$_{2}$ can be described
by a simple model based on a second-order differential equation (\ref{eq:accumulation})
\begin{equation}
\frac{d[\mbox{T}]}{dt}=2F\text{[T\ensuremath{_{2}}]}-K_{T}^{r}\text{[T]\ensuremath{^{2}},}\label{eq:accumulation}
\end{equation}
where {[}T{]} and {[}T$_{2}${]} are the concentrations of tritium
atoms and molecules, $F$ is the production rate (dissociation probability
per second) of T-atom pairs and $K_{T}^{r}$ is the temperature dependent
recombination constant. Both production rate and temperature dependent
recombination can be extracted by fitting the density growth rate,
$\frac{d[\mbox{T}]}{dt}$, with a square function and keeping $F$
and $K_{T}^{r}$ as fitting parameters. Using the value of $F$ extracted
from the fit to the data of Sample 2 we estimated that each electron
released after $\beta$-decay of T creates $\sim$50 unpaired H or
T atoms. This result is in a fair agreement with that of Collins at
al. \cite{Collins92} who found that the production effciency at 2.1$\,$K
is about 70 atoms per disintegration event \cite{Collins92}. However
this value is larger than that reported by Sharnoff and Pound: 22
unpaired atoms in their D$_{2}$ samples \cite{Sharnoff_Pound}.

\begin{figure}
\includegraphics{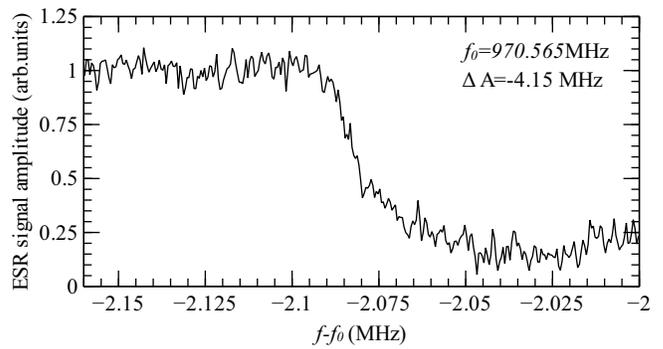}\protect\protect\caption{ENDOR of atomic tritium in a T$_{2}$ matrix\label{fig:ENDOR_TinT2}.
The calculated transition frequency for atoms in the gas phase is
denoted as $f_{0}$}
\end{figure}

Measuring the spectra of electron-nuclear double resonance (ENDOR)
in ``pure'' T$_{2}$ samples we found a clear transition at $\approx$2.085$\,$MHz
lower frequency than the value calculated for free atoms (see Fig.\ref{fig:ENDOR_TinT2}).
This corresponds to a negative change of the hyperfine constant, similar
to what has been observed for H in solid H$_{2}$ and D$_{2}$ matrices.
We carefully checked the frequency range a few MHz above the free
atom value, but no other transitions were detected there. This result
agrees with the previous results by Lambe \cite{Lambe60} and Sharnoff
and Pound \cite{Sharnoff_Pound} but contradicts to the data of Collins
et al. \cite{Collins92} who reported a large positive increase of
the hyperfine constant for T atoms in a T$_{2}$ matrix.

\subsection{Tritium:hydrogen mixtures}

We studied three T$_{2}$:H$_{2}$ mixture samples (Samples 4,5,6).
The maximum concentration of atomic hydrogen was achieved in the Sample
5: $1.1\times10^{20}$cm$^{-3}$. However only modest concentrations,
$\sim5\times10^{19}$ and $\simeq2.5\times10^{19}$cm$^{-3}$ were
achieved in the 4th and 6th samples (Tab.\ref{tab:A-summary-table}).
The T:H ratios in all three samples were much smaller than one would
expect from the ratio of T:H atoms in the gas mixture used for preparing
the samples. This gives evidence of a fast T-to-H conversion due to
the isotopic exchange reaction (\ref{eq:faster}).

\begin{figure}
\includegraphics{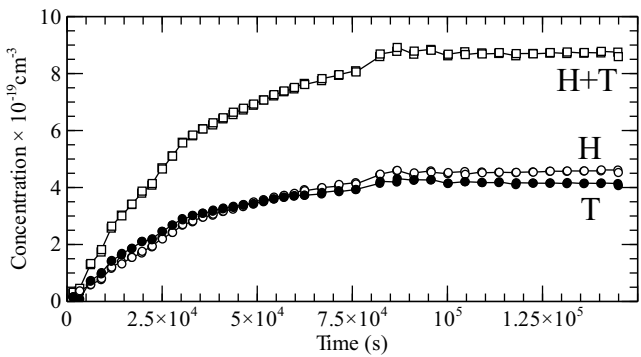} \protect\protect\caption{Time evolution of H and T concentrations in the sample 4 stored at
80$\,$mK\label{fig:T2_5H2}}
\end{figure}

The kinetics of the isotopic exchange reactions (\ref{eq:faster})
and (\ref{eq:slower}) was studied in the 80$\,$nm T$_{2}$:H$_{2}$
sample 4. The time evolution of H and T concentrations in this sample
is shown in Fig.\ref{fig:T2_5H2}. The H and T lines appeared few
minutes after deposition and had equal amplitudes. A nearly 1-to-1
ratio of their areas remained for the whole measurement and can be
explained by the very fast reaction (\ref{eq:faster}). A weak increase
of the H:T ratio after t=80000$\,$s while the total {[}H{]}+{[}T{]}
concentrations remained conserved can be explained by the contribution
from the reaction (\ref{eq:slower}). Based on that we concluded that
the reaction (\ref{eq:slower}) indeed proceeds much slower and we
can neglect its contribution to the {[}H{]} growth for our estimate
of the rate of the faster exchange reaction (\ref{eq:faster}).

The evolution of T and H atomic concentrations can be expressed by
the differential equations:

\begin{equation}
\frac{d[T]}{dt}=2F([\mbox{T}_{2}]+\frac{1}{2}[\mbox{HT}])-K^{ex}[\mbox{T}][\mbox{H}_{2}]-K_{T}^{r}[\mbox{T}]^{2}-K_{TH}^{r}[\mbox{T}][\mbox{H}]\label{eq:T reaction}
\end{equation}

\begin{equation}
\frac{d[H]}{dt}=2F([\mbox{H}_{2}]+\frac{1}{2}[\mbox{HT}])+K^{ex}[\mbox{T}][\mbox{H}_{2}]-K_{H}^{r}[\mbox{H}]^{2}-K_{TH}^{r}[\mbox{T}][\mbox{H}]\label{eq:H reaction}
\end{equation}
where $K^{ex}$ is a second-order rate constant of the exchange reaction
(\ref{eq:faster}) and $K^{r}$ are the recombination rate constants.
The production of H atoms includes two terms: dissociation of H$_{2}$
and HT molecules by $\beta$-particles and the isotopic exchange reaction
(\ref{eq:faster}). If we consider the initial part of the measurement
immediately after the deposition of the film, the concentrations of
atoms are close to zero and we have only the first terms in equations,
e.g. production due to the dissociation. Due to the much larger concentration
of T$_{2}$ in the sample, we should observe a proportionally larger
rate of the atomic {[}T{]} growth with respect to {[}H{]}. As one
can see from Fig.\ref{fig:T2_5H2}, both {[}T{]}(t) and {[}H{]}(t)
curves start with the same slope and the densities of T and H are
equal to each other during the whole measurement. This implies that
the exchange term in equations grows faster than we can observe with
our ESR technique, and the exchange reaction (\ref{eq:faster}) occurs
on the time scale $\tau^{ex}$< 100$\,$sec, a typical time interval
between sweeps of ESR lines. Having {[}H{]}={[}T{]} in the course
of measurement allows us to fit the rate of the H atom growth, $\frac{d[H]}{dt}$
with a parabolic function, $f(x)=k+k^{ex}x-k^{r}x^{2}$, where $x\equiv[T],[H]$,
$k=2F([\mbox{H}_{2}]+\frac{1}{2}[\mbox{HT}])$ is a constant representing
the rate of production, $k^{r}=K_{TH}^{r}+K_{H}^{r}$ is the effective
recombination constant, $k^{ex}=K^{ex}[\mbox{H\ensuremath{_{2}}}]$.
From the data of Fig.\ref{fig:T2_5H2} we extracted the values for
$K^{ex}$ and $k^{r}$ $3(2)\times10^{-26}$cm$^{3}$s$^{-1}$ and
10(5)$\times10^{-25}$cm$^{3}$s$^{-1}$, respectively. The relatively
large error in these rate constants is caused by a large uncertainty
in the concentration of H$_{2}$ in our mixtures. Such uncertainty
appears due to the small amount of tritium gas we worked with ($\sim$5$\,$
$\mu$mol) which created difficulties in preparing T$_{2}$:H$_{2}$
mixtures with a small admixtures of H$_{2}$. The exchange reaction
rate found here is smaller than that calculated by Aratono \cite{Aratono00}
($2\times10^{-25}$cm$^{3}$s$^{-1}$), while the recombination rate
agrees with that found in pure T$_{2}$ samples. The recombination
constant of tritium atoms obtained by Collins et al. \cite{Collins92}
is $K_{0}=1\times10^{-24}$cm$^{3}$s at 2.1$\,$K.

Estimating the rate of the slower exchange reaction (\ref{eq:slower})
requires a more sophisticated analysis because the differential equations
(\ref{eq:T reaction}) and (\ref{eq:H reaction}) will contain more
terms. Also a longer measurement might be required.

\begin{figure}
\includegraphics{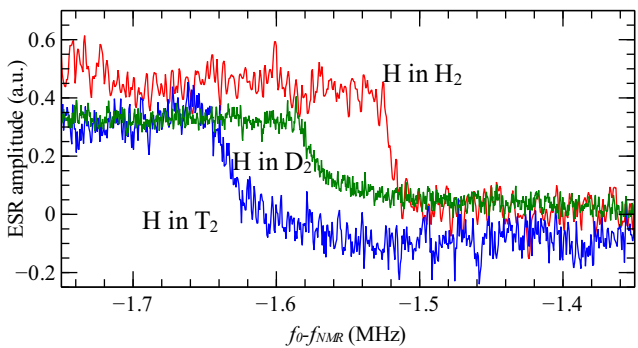} \protect\protect\caption{ENDORs of H atoms in matrices of hydrogen isotopes\label{fig:ENDORs-of-H}.
The calculated transition frequency for H atoms in the gas phase is
denoted as $f_{0}$}
\end{figure}

\section{Discussion}

We studied several samples of atomic tritium and hydrogen stabilized
in matrices of T$_{2}$ and T$_{2}$:H$_{2}$. The maximum concentrations
of H and T atoms were limited by their recombination. Recombination
of atomic hydrogen in solid hydrogen matrices at low temperatures
proceeds in two steps: a diffusion stage when atoms approach each
other by the distance of a lattice constant followed by rapid recombination
into molecules. Kumada\cite{Kumada} showed that diffusion of hydrogen
atoms at temperatures $\simeq1\,$K proceeds in a series of exchange
tunneling reactions 
\begin{equation}
\mbox{H+H\ensuremath{_{2}\rightarrow\mbox{H\ensuremath{_{2}+\mbox{H}}}}}\label{eq:Hdiff}
\end{equation}
Similar exchange reaction also governs diffusion of atomic deuterium
in solid D$_{2}$, and is also expected in tritium. However its rate
should be much smaller than those for hydrogen and deuterium due to
a larger mass and smaller zero-point energies of the reactants. The
rate of the gas phase reaction (\ref{eq:Hdiff}) at 4.2$\,$K, $k\sim10^{-25}$cm$^{3}$s$^{-1}$,
was calculated by Takayanagi et al.\cite{Takayanagi90} and is in
a fair agreement with the recombination rates of H atoms in solid
H$_{2}$. The measured recombination rates for D atoms in D$_{2}$
are about 2 orders of magnitude smaller than those for hydrogen\cite{Iskovskikh86}.
Based on that, we may expect that the recombination rates of T atoms
in pure T$_{2}$ at the same temperatures should be at least 1-2 orders
of magnitude smaller than those for D in D$_{2}$.

The reaction (\ref{eq:Hdiff}) and its isotopic analogs have a large
activation barrier, $E_{a}\simeq4600\,$K, and proceed at low temperatures
by tunneling. Any impurity or crystal defect can perturb the periodic
potential of the matrix and create an energy level mismatch for a
tunneling event. The mismatch can be compensated by phonons. Ahokas
et al.\cite{Ahokas10} reported enhancement of recombination rate
during recombination of gas-phase H atoms at the surface of their
H$_{2}$ samples. The recombination rates of T atoms in our samples,
$k\sim10^{-24}-10^{-25}$cm$^{3}$s$^{-1}$, are much larger than
what could be expected from the rates of the isotopic exchange reaction
$\mbox{T+T\ensuremath{_{2}}\ensuremath{\rightarrow}T\ensuremath{_{2}+\mbox{T}}}$
which should have a rate at least 3 orders of magnitude smaller than
the value we obtained. This is an upper limit estimate for this exchange
reaction in the abscence of any limiting factors such as energy level
mismatch due to crystal defects. An alternative way can be a physical
diffusion of T atoms related to formation of vacancies. Gaines et
al. \cite{Gaines89} estimated the activation energy for physical
diffusion of T in T$_{2}$ as 411$\,$K which is about 2 times larger
than that for H in H$_{2}$ (195$\,$K)\cite{Leachphd72}. Based on
that, one may expect a cross-over temperature from quantum diffusion
to Arrhenius-like behavior to be about 2 times larger for tritium
than that for hydrogen: about 9$\,$K against 4.5$\,$K. The pre-exponential
factors for these processes were found to differ only by a factor
of 6 \cite{Gaines89}, \cite{Leachphd72} which does not influence
the result.

Much higher concentrations of T atoms in ``pure'' T$_{2}$ Sample
2 compared to T$_{2}$ samples with H$_{2}$ admixtures (4,5,6) leads
us to expect a faster recombination of H atoms compared to T. Analysing
the differential equations for the evolution of {[}H{]} and {[}T{]}
concentrations in the T$_{2}$ sample 3 and T$_{2}$: H$_{2}$ sample
6 we obtained a 3 times larger recombination rate for H atoms in the
latter sample.

$\beta$-decay of tritium results in the formation of a large number
of non-equlibrium vacancies and phonons which may lead to a significant
enhancement of physical diffusion. Ebner and Sung \cite{Ebner_Sung}
considered two cases of diffusion through vacancies in H$_{2}$: tunneling
and hopping to the empty sites. Pure tunneling of a T or H atom in
a T$_{2}$ matrix should be suppressed due to a large number of lattice
defects but it can be enforced by phonons generated by tritium decay.
The phonons may not only help to compensate for the energy level mismatch
but also help atoms to hop over the barrier and stimulate physical
diffusion followed by recombination. Due to the very large energy
released in the tritium decay the real temperature of phonons may
be much higher than the typical storage temperatures $\sim100\,$
mK used in our experiments. Recombination of atoms brings extra heating,
which leads to enhancement of diffusion and recombination rate at
higher densities of atoms. Such positive feedback in the system finally
leads to a thermal explosion of atoms which occurs at a certain critical
density. At high enough concentration of atoms even a small increase
of temperature may cause an instantaneous increase of number of phonons
and provoke a so-called stimulated explosion.

The fact that only a fraction of atoms in the film recombines during
explosions may probably be explained by better stability of atoms
more close to the film substrate due to better cooling whereas the
atoms closer to the top of the films recombine preferentially. This
also explains the absence of thermal explosions in thin films and
their suppression after adding helium to the sample cell which provides
better cooling to the upper surface of the film.

We tried to estimate possible overheating of the tritium film during
a recombination explosion using a simple heat balance model. We assume
that the heat released in a recombination event is conducted by the
phonons from the T$_{2}$ film to the quartz microbalance and further
down into the SC lower volume, where part of the recombination energy
is removed by sublimation of tritium molecules. From the frequency
change of our QM we know that $\approx$5\% of the T$_{2}$ film becomes
sublimated after typical explosive recombination when the density
of atoms is decreased by $4\times10^{19}$ cm$^{-3}$ (see Fig.\ref{fig:Explosions})
corresponding to recombination of $\sim10^{15}$ atoms. Taking the
sublimation heat of solid T$_{2}$, H=1400$\,$J/mol\cite{souers1986hydrogen},
we evaluate that about 1/3 of energy released in recombination of
atoms is removed from the sample by the sublimated molecules. The
rest of the energy is conducted into the lower volume of the SC and
passes through several boundaries between solid hydrogen and gold
electrode, gold-quartz and finally gold-helium interfaces. It turns
out that the boundary thermal resistance between gold electrode and
T$_{2}$ film becomes the bottleneck for cooling the sample. The boundary
thermal resistances for gold-quartz and solid T$_{2}$-gold interfaces
were calculated using the acoustic mismatch model, which agreed with
a factor of 2 with experimental values \cite{Reynolds76}. The temperatures
for all layers of the microbalance: both gold electrodes and the quartz
were estimated by a simple model, similar to the estimate by Wyatt\cite{Wyatt92}
using the equations:

\[
\dot{q}=G_{ij}A(T_{h}^{n}-T_{c}^{n})
\]

Here $G_{ij}$ is the thermal boundary conductance between interfaces
$i$ and $j$ (e.g. solid hydrogen and gold), $A$ is the QM surface
area (1$\,$cm$^{2}$), $n$=4 for the solid-solid interfaces and
$n$=5 for the interfaces between liquid helium and solids. T$_{h}$
and T$_{c}$ are the temperatures of the hotter and colder layers
respectively. For the estimate of the heat power $\dot{q}$ we need
to know the time duration of the recombination explosion event. From
our bolometer response we obtain the value of $\tau_{rec}\lesssim100\,m$s,
while the light flashes observed by Mapoles et. al.\cite{Mapoles90}
indicate that it can be shorter than 1$\,m$s. Using the heat conductance
model above for the duration of the recombination explosion of 1 $m$s
we evaluate overheating of the tritium film to $\sim$16$\,$K, and
$\sim$12$\,$K if the explosion occurs during 10 $m$s. Both temperatures
are high enough to evaporate solid tritium, and for substantial enhancement
of the rate of the physical diffusion of T atoms.

Analysis of our data from ENDOR measurements allows making certain
conclusions about the structure of the tritium films we studied. The
lattice structure is estimated from the influence of the host molecules
on the electron clouds of unpaired atoms. Depending on a lattice site
either a long-range attractive van-der-Waals interaction or a short-range
Pauli repulsion prevails. A negative change of the hyperfine constant
corresponds to the substitutional lattice sites where the attractive
van-der-Waals contribution takes over. In this work we observed a
single ENDOR transition for T atoms in solid T$_{2}$ corresponding
to a $\approx-4.15\,$MHz shift of the hyperfine constant with respect
to the free atom transition. We have also performed ENDOR measurements
for H atoms in a T$_{2}$ matrix. We present it in Fig.\ref{fig:ENDORs-of-H}
together with the spectra in the matrices of H$_{2}$ and D$_{2}$
for the sake of comparison. The change of the hyperfine constant of
H atoms stabilized in a T$_{2}$ matrix was found to be -3.17$\,$MHz
which is about 0.1$\,$MHz larger compared to that of H atoms in the
matrices of other hydrogen isotopes. Tritium molecules have the lowest
zero-point energy among the hydrogens and so the lattice constant
for solid T$_{2}$ is the smallest as compared to the other isotopes.
The H atoms placed in more quantum matrices of lighter hydrogens are
pushed further away by the molecules and the hyperfine constant is
closer to that of free atoms. This also explains why the shift of
the ENDOR transition of H in T$_{2}$ is larger than in H$_{2}$.
Observation of a single ENDOR transition in all four cases indicates
that the impurity atoms occupy the same substitutional site in the
lattice, and the lattice type is most likely the same for all matrices.

\section{Conclusions}

In conclusion, we reported on the first ESR study of T and H atoms
stabilized in ``pure'' T$_{2}$ and T$_{2}:$H$_{2}$ matrices at
temperatures down to 70$\,$mK. The concentrations of T atoms approaching
$2\times10^{20}$cm$^{-3}$were reached in pure T$_{2}$ films. The
record-high concentrations of H atoms $\mbox{\ensuremath{\sim}}1\times10^{20}$cm$^{-3}$
were reached in T$_{2}$:H$_{2}$ solid mixtures. It turned out that
the isotopic exchange reaction T+H$_{2}\rightarrow$HT+H proceeds
with a much higher efficiency compared to reaction T+HT$\rightarrow$T$_{2}$+H
and results in a spectacular T-to-H conversion in T$_{2}$:H$_{2}$
mixtures. The accumulation of H and T atoms was limited by their recombination
which also occurred in an explosive manner depending on the storage
conditions. We suggest that the main mechanism for H and T migration
in solid T$_{2}$ is physical diffusion related to tunneling or hopping
to vacant sites in the lattice in contrast to isotopic chemical reactions
which govern diffusion of H and D atoms created in H$_{2}$ and D$_{2}$
matrices by other methods. 
\begin{acknowledgments}
We acknowledge funding from the Wihuri Foundation and the Academy
of Finland grants No. 258074, 260531 and 268745. This work is also
supported by NSF grant No DMR 1209255. S.S. thanks UTUGS for support.
\end{acknowledgments}

 \bibliographystyle{apsrev}
\bibliography{T_paper_21_09_Arxiv_submission}

\end{document}